\documentclass[superscriptaddress,twocolumn]{revtex4-1}

\usepackage[utf8]{inputenc}

\usepackage{float}

\usepackage{braket}

\usepackage{xcolor}

\usepackage{indentfirst}

\usepackage{tikz}
\usetikzlibrary{calc}

\newcommand{\ire}[1]{{ #1}}

\usepackage{amsmath}
\usepackage{latexsym}
\usepackage{amsfonts}
\usepackage{amssymb}

\usepackage{comment}

\usepackage[colorlinks=true,linkcolor=blue,urlcolor=blue,citecolor=blue,pdfusetitle]{hyperref}

\usepackage{ragged2e}
\justifying

\begin{document}

\title{Work statistics and Entanglement across the fermionic superfluid-insulator transition}
\author{Krissia  Zawadzki}
\affiliation{School of Physics, Trinity College Dublin,
 College Green, Dublin 2, Ireland}

\author{Guilherme A. Canella}
\affiliation{S\~ao Paulo State University, 14800-090, Araraquara, S\~ao Paulo, Brazil}
\author{Vivian V. Fran\c ca}
\affiliation{S\~ao Paulo State University, 14800-090, Araraquara, S\~ao Paulo, Brazil}

\author{Irene D'Amico*}
\affiliation{Department of Physics, University of York, York YO10 5DD, UK}

\begin{abstract}
    \ire{Entanglement in many-body systems may display interesting signatures of quantum phase transitions and similar properties are starting to be encountered in the analysis of work fluctuations. Here, we consider the fermionic superfluid-to-insulator transition (SIT) and relate its entanglement properties with its work distribution statistics. The SIT is modeled by the attractive fermionic Hubbard model in the presence of randomly distributed impurities. The work distribution is calculated across two quench protocols, both triggering the SIT. In the first, the concentration of impurities is increased;  in the second, the impurities' disorder strength is varied. Our results indicate that, the critical state that induces minimization of the entanglement also maximizes the  average work. We demonstrate that, for this state, density fluctuations vanish at all orders, hence  all central moments of the work probability distribution are exactly zero at criticality.  For systems undergoing a precursor to the transition (short chains with finite impurity potential) numerical results confirm these predictions, with higher moments further from the ideal result.   For both protocols, at criticality, the system absorbs the most energy with almost no penalty in terms of fluctuations:  ultimately this feature could be used to implement a quantum critical battery. The effects of temperature on these signatures of critical behaviour  are also investigated and shown to favor work extraction for high enough temperatures.}


\end{abstract}

\maketitle

\section{Introduction}

The behavior of the work that can be extracted or absorbed by a many-body quantum system is an important question in quantum thermodynamics.
In interacting many-body quantum systems, correlations that are inherently quantum may manifest in the form of entanglement, and may show interesting properties at criticality, including universal scaling \cite{Osterloh2002scaling}. This has inspired a series of investigations on the work statistics across a quantum phase transition (QPT)\cite{Adolfo-PhysRevLett.101.120603,Mascarenhas-PhysRevE.89.062103,Work_statistics_QPT-PhysRevLett.124.170603, Landi-PhysRevResearch.2.033279,Steve-critical-PhysRevB.94.184403}.
Many previous studies have focused on characterizing the scaling properties of the work statistics following a sudden quench across the QPT \cite{Work_fluctuations_MB-PhysRevX.4.031029, QTD-critical-PhysRevB.93.201106,Work_statistics_QPT-PhysRevLett.124.170603, Steve-critical-PhysRevB.94.184403,Steve-excited-stateQPT-PhysRevE.103.032145}. Recently, the effects of the finite-time dynamics have started to be addressed, with special attention to the statistics beyond the second moment \cite{Slow_drive_work_fluctuations-PhysRevLett.123.230603,QWKrissia,Work-finite-t-PhysRevE.104.L062102,Miller-work-finite-t-PhysRevLett.125.160602}.
These studies opened the way for novel applications, such as heat engines implemented using many-body systems as working medium,  \cite{isolated-heat_engine-bosons-PhysRevLett.108.085303, Spin-heat-engine,Rydberg-Atom-Engine,critical-heat-engine,heat-engine-MBL-PhysRevB.99.024203,Ising-heat-engine-piccitto2022ising}, which can be regarded as an example of quantum advantage when exploiting quantum correlations \cite{Herrera2023correlationboosted}. It has been demonstrated that interactions between particles may allow for a boost in their efficiency, which is much larger than that of a non-interacting system of the same size \cite{heat-engine-efficiency, Quantum_supremacy_MB_heat_engines_Jaramillo_2016,Herrera2023correlationboosted}.
In the particular case of a working medium undergoing a QPT, criticality may lead to supralinear scaling of power \cite{critical-heat-engine,Campisi2016power}, however, even at the level of simple models, the practical implementation of these engines requires special control techniques to operate thermal cycles at finite-time, in addition to minimizing fluctuations of work and output power \cite{Adolfo_HE_STA,Victor_MB_HE_STA, Marti_optimal_HE_critical}.

 Entanglement has emerged as a powerful tool to detect and characterize QPTs \cite{Osterloh2002scaling, QuantumEntanglement,Classical-quantum}. Pioneer works exploring  bi-partite entanglement in critical models  showed that  QPTs {may be} associated to an entanglement extremum or to the entanglement non-analytical behaviour \cite{Osterloh2002scaling,Osborne-PhysRevA.66.032110,Wu-2006,Vivian-2013-entangle}. Following these ideas, various measures of entanglement have been employed as witnesses of changes in phases of matter. In the context of localization in the presence of disorder, metals \cite{Entanglement-localization,Entanglement-Anderson,disordered-latticeDFT,entanglement-QPT}, bosonic systems \cite{measure-entanglement,LAFLORENCIE,entanglement-SIT-bosons,SIT-Bosons-PhysRevLett.92.130403}, spinless \cite{area-law-violation,momentum-space-entanglement} and spinfull  \cite{Canella2020-entanglement,Canella2021-MottAnderson,Canella2022-MottAnderson} fermions were explored. The fermionic superfluid to insulator transition (SIT) \cite{Anderson-diffusion,ANDERSON-dirty-SC,Bloch_Greiner_SIT} is one of the QPTs displaying interesting entanglement properties that has been least explored in applications to quantum thermodynamics. The SIT in disordered systems is a paradigmatic problem in quantum physics, as it spurs from the competition between free and coherent mobility, and localization of particles.
Advances in experiments with optical lattices and ultracold atoms have made it possible to quantum-simulate the SIT with ultracold bosons \cite{Greiner2002quantum} and fermions \cite{Chin2006superfluid_fermions}.  For strongly correlated bosons in a Tonks-Girardeau gas, modeled by the Bose-Hubbard model, it has been shown that the superfluid-insulating transition is accompanied by a pinning QPT\cite{Haller2010pinning}, which can be achieved at infinitesimally weak lattice potentials \cite{Pinning-PhysRevA.86.033620}.
In fermionic systems, superfluidity can emerge from attractive Coulomb interactions \cite{Giamarchi_book_1D,Rosch}, while an insulating behavior is due to repulsion \cite{Esslinger2010}.
In the presence of impurities, the interplay between attractive interactions and disorder leads to a richer phase diagram, in which an insulating, superfluid, and localized behavior can be present. In fermionic systems  the conditions under which the SIT occurs are still under investigation.  Refs. \cite{Vivian-2017-superfluid,Canella2019superfluid} investigated the average single-site entanglement in disordered chains with attractive Coulomb electron-electron interactions. These results indicated the absence of a critical potential $V$ for the emergence of the SIT triggered by disorder, whilst the existence of a critical concentration $C_C$ and a critical particle density for which the entanglement is minimum. These recent results could help fine tuning the parameters controlling the disorder landscape and attractive couplings to induce the fermionic SIT in, e.g., optical lattices.

Motivated by the prospects of exploiting the SIT in thermodynamic cycles \cite{critical-heat-engine} and by its interesting entanglement structure in fermionic systems \cite{Canella2019superfluid},
we compare the average work statistics of two types of sudden quenches of interest to the SIT. In the first  protocol, the concentration of impurities is changed by adding an extra impurity to the initial state. In the second protocol,  we consider the situation in which the impurity potential strength is instantaneously increased. For a system initially prepared at temperature $T=0$, we show that if the initial ground-state has minimum entanglement,  the average work will be maximum at the critical concentration $C_C$, whilst its variance will be minimized at the same rate in which fluctuations in density-density correlations decay. We also inspect the skewness of the work distribution, which provides a measure of non-Gaussianity  and has been shown to be very sensitive to phase transitions for systems driven in finite time\cite{QWKrissia,Zawadzki_nonGaussian}.
 We observe a similar sensitivity for the SIT, with the skewness providing more information than the variance about the subtle interplay between impurity strength and Coulomb interaction in the second protocol.
 Finally, the effects of temperature are also discussed. As the temperature increases, features of the quantum phase transition are suppressed, and we observe that thermal fluctuations favor work extraction (instead of absorption) whenever thermal excitations become larger than the typical energy scales of the system.

\section{Modeling the superfluid-insulator transition in fermionic systems}

The superfluid-insulator transition in fermionic systems can be described by the Hubbard model with attractive electron-electron interactions in the presence of disorder \cite{Canella2019superfluid}. In a 1D chain of $L$ sites, the Hamiltonian reads

\begin{equation}
    \hat{H} = -J\sum^{L}_{i,\sigma}\left(\hat{c}^{\dagger}_{i,\sigma}\hat{c}_{i+1,\sigma} + \hat{c}^{\dagger}_{i+1,\sigma}\hat{c}_{i,\sigma}\right) + U\sum^{L}_{i}\hat{n}_{i,\uparrow}\hat{n}_{i,\downarrow} + \sum_i^L V_i\hat{n}_i,
\end{equation}
where \ire{$J>0$} is the hopping parameter, $U<0$ is the attractive on-site Coulomb interaction and $V_i$ is the strength of the impurity potential at site $i$, the local disorder. The number operator is $\hat{n}_{i,\sigma} = \hat{c}^{\dagger}_{i,\sigma}\hat{c}_{i,\sigma}$ where $\hat{c}^{\dagger}_{i,\sigma}$ is the creation operator for a fermion with $z$-spin component $\sigma = \uparrow,\downarrow$ in the $i$-th site, such that $n=n_\uparrow+n_\downarrow$ and $m=n_\uparrow-n_\downarrow$ are particle density and magnetization, respectively, with $n_\sigma = \sum_i \text{Tr}[  \hat{n}_{i,\sigma}  \hat{\rho} ] $ with $\hat{\rho}$ the state of the system.

For $L\to\infty$ and in the absence of disorder ($V_i=0$),  increasing attractive interaction drives the system from a superfluid state of  weakly-coupled pairs (BCS-like) to a phase of tightly coupled dimers (Bose-Einstein limit). In contrast, by triggering the local disorder $V_i$ at a fixed interaction $U$, the pairs tend to localize and the system undergoes a superfluid to insulator transition. Localization can be reached in a superfluid by either increasing the disorder intensity or lowering the particle density \cite{Canella2019superfluid}.

It has been recently shown\cite{Canella2019superfluid} that, by considering \ire{$N_i$} point-like impurities of same strength $V_i=V$, randomly distributed along the $L$-size chain,  the level of localization in the SIT depends on the concentration $C = (N_i/L)\times 100\%$. For sufficiently strong disorder strength $|V| \gg |U|$, it is possible to achieve full localization at the critical concentration $C_C = 100n/2$. For $|V|\rightarrow \infty$, this full localization is characterized by real-space pairs localization, marked by zero entanglement \cite{Canella2019superfluid}.

Here we consider the same disorder landscape $-$ a certain concentration $C$ of pointlike impurities randomly placed along the chain with disorder strength $V$ $-$. Both long ($L=100$) and short ($L = 8$) superfluid chains with $m=0$ are considered. To avoid features related to specific configurations, for all quantities we perform an average over  100 samples for the long chain case, and over all possible impurity configurations for $L=8$ chains.

For $L=100$, we consider the ground-state average single-site  entanglement, which has been successfully applied to explore quantum phase transitions in the Hubbard model \cite{Canella2019superfluid,Vivian-2013-entangle,Canella2020-entanglement,Canella2021-MottAnderson,Canella2022-MottAnderson,MetricTatiPicoli}. We follow the same approach of Refs. \cite{Canella2019superfluid,Canella2020-entanglement,Canella2021-MottAnderson,Canella2022-MottAnderson} obtaining the ground-state single-site entanglement via the linear entropy. This is averaged first over the $L$ sites, ${\mathcal L}=(1/L)\sum_i (1-Tr[\rho_i^2])$, where $\rho_i$ is the reduced density matrix of site $i$,  and then over 100 samples of different impurity configurations compatible with each of the protocols,   $\bar{\mathcal L}=(1/M)\sum_m {\mathcal L}_m$.

\subsection{Statistics of work in a sudden quench}

One can perform (or extract) work on an isolate quantum system by changing the parameters $\vec{g}$ in the time-dependent Hamiltonian $\hat{H}({\vec{g}}_t)$ taking the system out-of-equilibrium. In this process, all possible transitions between the eigenestates of the initial $\hat{H}(\vec{g}_0)$ and final $\hat{H}(\vec{g}_f)$ Hamiltonians may be involved, determining the change in energy as well as its fluctuations. Experimentally, the quantum work is accessible by means of spectroscopic methods \cite{EmergentTherm,CharacFun,Marcela_PRL2021}, and the work distribution has been reconstructed in a liquid-state nuclear magnetic resonance platform using small molecules as working fluid \cite{Batalhao,Marcela_PRL2021}.

In a sudden quench, the transitions are instantaneous and the system does not have time to adapt. If the system is prepared in a superposition of the eigenstates $\ket{n^{0}}$ of $\hat{H}(\vec{g}_0)$ with weights $p_n$, the work probability distribution $P(W)$ is calculated as follows \cite{Lutz_work_distribution}

\begin{equation}
    P(W) = \sum_{n,m}\delta\left[ W - \left(\epsilon_m^{0^+} - \epsilon_n^{0}\right)\right]p_n p_{m|n},
\end{equation}
where $\{\epsilon_n^{0}\}$ and $\{\epsilon_m^{0^+}\}$ are the eigenvalues of  $\hat{H}(\vec{g}_0)$ and $\hat{H}(\vec{g}_f)$, respectively, and $p_{m|n} = |\braket{m^{0^+}|n^{0}}|^2$ is the probability to find the system in the $m$-th eigenstate $\ket{m^{0^+}}$ of $\hat{H}(\vec{g}_f)$ at $t>0^+$ given the dynamics started in state $\ket{n^{0}}$ at $t=0$.

Associated with the work distribution, we have the $k$-th order central moments $\braket{(W-\braket{W})^k}$ which, for $k>1$ can be calculated recursively, starting from the average work

\begin{equation}
   \braket{W} = \sum_{n,m} \left(\epsilon_m^{0^+} - \epsilon_n^{0}\right) p_n p_{m|n}.
\end{equation}
\ire{Average work performed on the system correspond to $\braket{W}>0$; while extracted work is signalled by  $\braket{W}<0$.}
The second moment, $k = 2$, is the variance, associated with the energy fluctuations. The third moment quantifies the skewness ($k = 3$) of the work distribution, which is  related to the deviation from Gaussianity.

\subsection{Sudden quench protocols across the SIT}
\begin{figure*}[htb!]
     \centering
     \includegraphics[width=\textwidth]{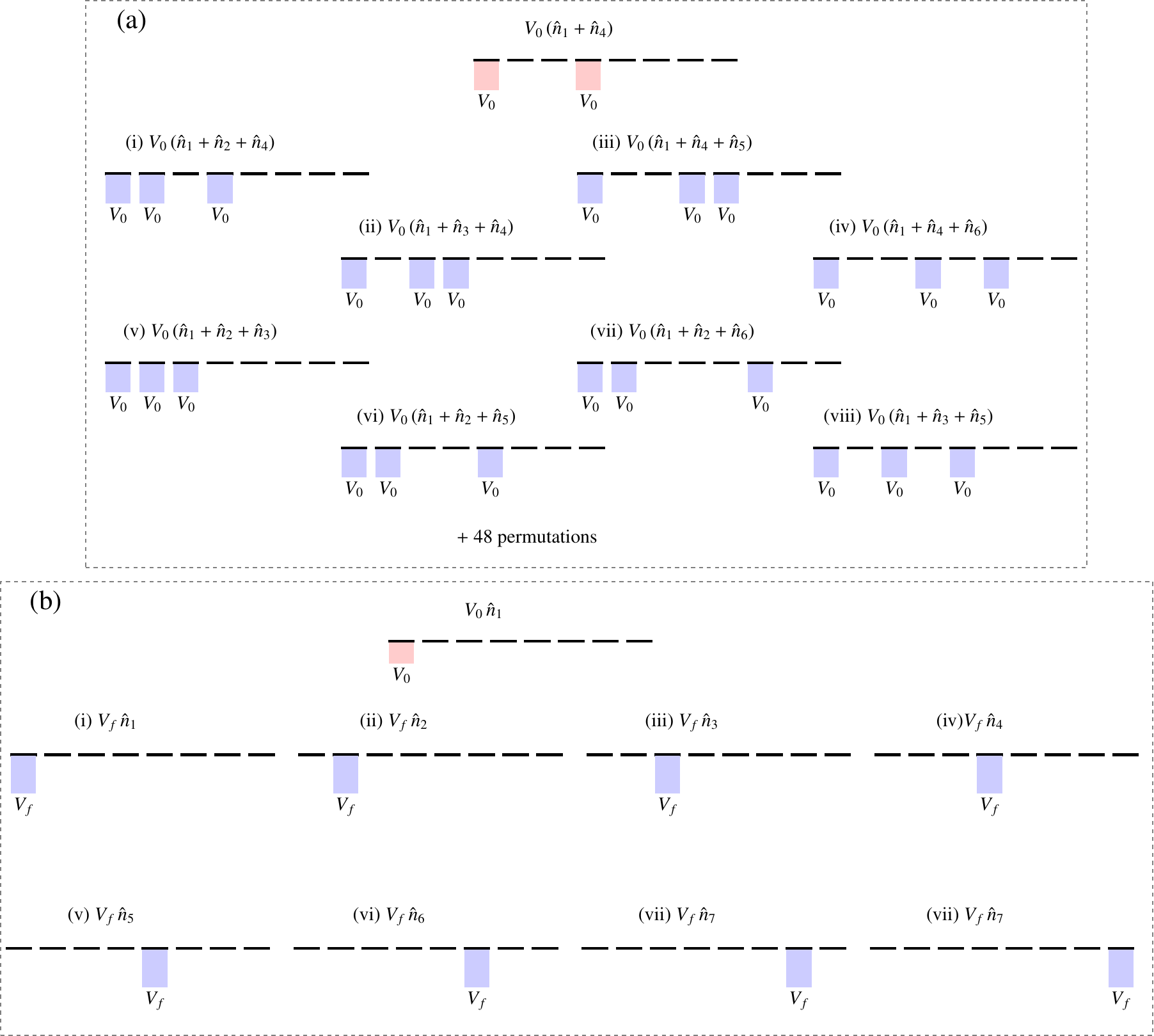}
     \caption{Illustration of sudden quench protocols across the superfluid-insulator transition in a chain of $L=8$ sites. Panel (a) shows a quench in the concentration of impurities and a few possible configurations starting from one of the 28 possible configurations with $N_i =2 $ impurities. For 3 impurities (final state), there are a total of 56 configurations and we show only 8 of them. Panel (b) shows a possible initial state for a quench in the disorder strength at fixed concentration with all possible final configurations with $N_i =1$ impurity.}
         \label{fig:protocols}
 \end{figure*}

\ire{We will study two sudden-quench protocols  across the SIT and the associated average work.}
In the first protocol we vary the impurity concentration $C$ for fixed values of $V$. For each initial concentration,  we consider in turn each of the possible corresponding impurity configurations.  For each initial impurity configuration, we consider in turn all possible configurations ($L=8$) or 100 samples of different impurity configurations ($L=100$) that can be achieved by adding one extra impurity, see Fig. \ref{fig:protocols}(a). For each initial and final configuration, we calculate the \ire{average} work and the moments of its distribution. For each initial concentration, we then average these quantities over all possible couples of initial and final configurations.

In the second protocol
initial and final Hamiltonians will have the same number of impurities but their position may vary and the final impurity potential $V_f$ will be in modulus larger than the initial potential $V_i$, see Fig. \ref{fig:protocols}(b). The results will be averaged over all possible initial and final configuration with same impurity concentration $C$.

For both protocols, in the sudden quench limit, all central moments can be expressed in terms of correlation functions of the density, as the terms of the Hamiltonian associated with the kinetic energy and the Coulomb repulsion remains the same.  Here we discuss in details the first three central moments.
The average work is a functional of the local densities of the initial state:
\begin{align}
    \label{eq:average_work}
    \braket{W} & = \text{Tr}[ ( \hat{H}_f - \hat{H}_0) \hat{\rho_0} ]  \\
    & = \sum_{j=1}^L \Delta v_{j}
 \text{Tr}[  \hat{n}_j  \hat{\rho_0} ] ,
 \label{eq:average_work+GS}
\end{align}
where $\Delta v_j =  (V_j^f - V_j^0 )$

The fluctuations of work, corresponding to the second central moment, depend on density-density fluctuations, as
\begin{align}
     \sigma_W^2 & = (\braket{W^2} - \braket{W}^2) \\
     & = \sum_j \sum_\ell \Delta v_j \Delta v_\ell
     (\text{Tr}[  \hat{n}_j \hat{n}_\ell  \hat{\rho_0} ]-\text{Tr}[  \hat{n}_j  \hat{\rho_0} ]\text{Tr}[\hat{n}_\ell  \hat{\rho_0} ])
\label{fluct_sigma}\end{align}
where we have used
\begin{align}
    \label{eq:MomentW2}
    \braket{W^2} & =  \text{Tr}[ ( H_f - H_0)^2 \hat{\rho_0} ] \\
    & =
    \sum_j \sum_\ell \Delta v_j \Delta v_\ell \text{Tr}[  \hat{n}_j \hat{n}_\ell  \hat{\rho_0} ] . \label{eq:MomentW2_GS}
\end{align}
Finally, the third central moment, the skewness, can be expressed  as the sum of densities' correlations

\begin{align}
     \mu_3 & = (\braket{W^3} - 3 \braket{W} \braket{W^2}
     + 2 \braket{W}^3) \label{newIDA}
     \\
     & = \sum_j \sum_\ell \sum_m \Delta v_j \Delta v_\ell \Delta v_m
     (\text{Tr}[  \hat{n}_j \hat{n}_\ell  \hat{n}_m\hat{\rho_0} ]
     \nonumber \\
     &
     -3\text{Tr}[  \hat{n}_j \hat{n}_\ell  \hat{\rho_0} ]\text{Tr}[  \hat{n}_m  \hat{\rho_0} ]
     + 2 \text{Tr}[  \hat{n}_j  \hat{\rho_0} ]\text{Tr}[\hat{n}_\ell  \hat{\rho_0} ]\text{Tr}[\hat{n}_m  \hat{\rho_0} ]),
\label{fluct_mu}\end{align}
where we have used
\begin{align}
    \label{eq:MomentW3}
    \braket{W^3} &= \text{Tr}[ ( H_f - H_0)^3 \hat{\rho_0} ] \\
    & =
    \sum_j \sum_\ell \sum_m \Delta v_j \Delta v_\ell \Delta v_m \text{Tr}[ \hat{n}_j \hat{n}_\ell \hat{n}_m \hat{\rho_0}].
    \label{eq:MomentW3_GS}
\end{align}

In particular, for the ground-state $| \Psi_0\rangle$ equations \ref{eq:average_work+GS}, \ref{eq:MomentW2_GS}, and \ref{eq:MomentW3_GS} are easily computed from the
density correlations  $\braket{\Psi_0 | \hat{n}_j | \Psi_0}$, $\braket{\Psi_0 | \hat{n}_j \hat{n}_\ell |\Psi_0}$, and $\braket{\Psi_0 | \hat{n}_j \hat{n}_\ell \hat{n}_i |\Psi_0}$.

The value of $\braket{W}$, $ \sigma_W^2$ and $ \mu_3$ will vary depending on the initial (and final) impurity configurations  of $H_0$ (of $H_f$). We will then consider their average over all possible configurations $N_c$ compatible with the relevant initial and final impurity concentrations ($L=8$). For $L=100$, $N_c$ will comprise 100 configurations. These averages are defined  as
\begin{align}
    \label{eq:average_averages_moments}
    \overline{\braket{W}} & = \frac{1}{N_c}\sum_i ^{N_c} \braket{W}_i \\
\overline{\sigma_W^2} & = \frac{1}{N_c}\sum_i ^{N_c} \braket{\sigma_W^2}_i \\
\overline{\mu_3} & = \frac{1}{N_c}\sum_i ^{N_c} \braket{\mu_3}_i.
\end{align}

While $\overline{\braket{W}}$ is the average of the average work, for the sake of simplicity, in the rest of the paper, we will refer to it simply as  the `average work'.

\section{Results}

\ire{
\subsection{The critical state } \label{anal}
Vanishing entanglement at criticality reveals that, for each configuration, the ground-state is localized at the impurity sites, each of which is exactly doubly occupied in the limit $|V|\gg |U|$ and $T=0$. In the occupation basis, this state is factorized and can be written as
\begin{align}
|\Psi_C\rangle = \otimes_{i} |j\rangle_i
\label{Psi_C}
\end{align}
with $j=\uparrow\downarrow$ if site $i$ contains an impurity and $j=0$ otherwise.

For this type of localized state, density fluctuations vanish at all orders that is
\begin{equation}
 \braket{\Psi_C | \otimes_{i}\hat{n}_i |\Psi_C}=\Pi_{i}\braket{\Psi_C | \hat{n}_i|\Psi_C}, \label{dens_fluct}
\end{equation}
where $i$ extends to any subset of sites.
By inspection of eqs. (\ref{fluct_sigma}), and (\ref{fluct_mu})
it is then clear that variance and skewness vanish exactly at criticality. As all higher central moments can be similarly written in term of density fluctuations, it follows from eq. (\ref{dens_fluct}) that all central moments of the work probability distribution exactly vanish at the critical state.}

\subsection{Quench in $C$: adding an extra impurity }
We start our analysis with the average work - eqs. \ref{eq:average_work} and \ref{eq:average_averages_moments}- at $T=0$ and the first protocol.

\begin{figure*}
    \centering
    \begin{tikzpicture}
        \node[label={[font=\normalsize, shift={(-0.5cm,-5mm), align=left}]above left:(a)}] (figa) at (0,0)
        {\includegraphics[width=0.61\columnwidth]{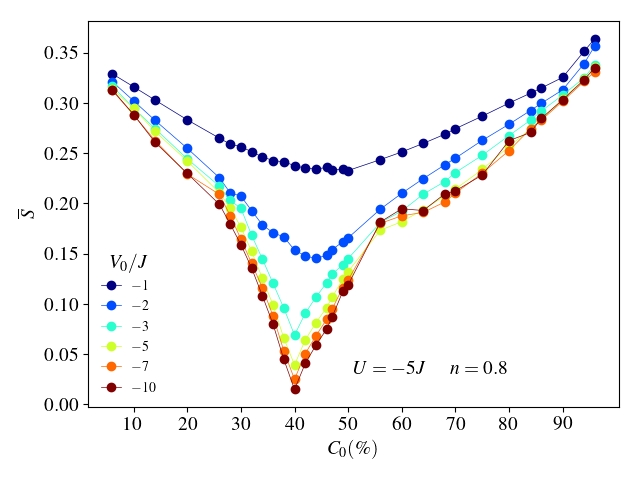}};

        \node[anchor=north west,
       label={[font=\normalsize,
       shift={(-0.3cm,-5mm), align=above }]above left:(b) }
        ] (figb) at ($(figa.north east)+(0.2cm,0)$)
        {\includegraphics[width=0.61\columnwidth]{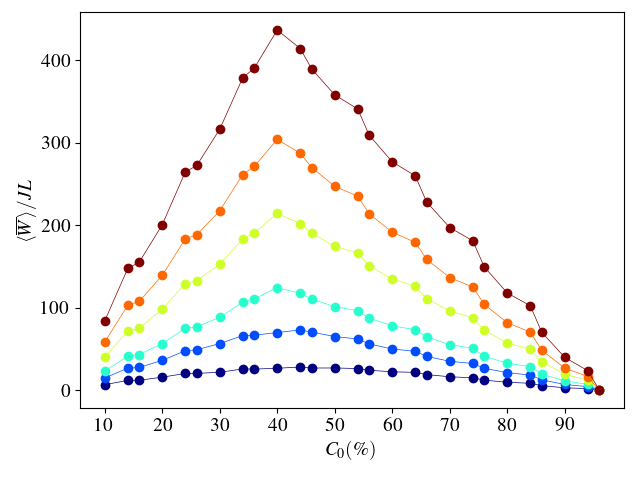}};

        \node[anchor=north west,
       label={[font=\normalsize,
       shift={(-0.3cm,-5mm), align=above }]above left:(c) }
        ] (figc) at ($(figb.north east)+(0.2cm,0)$)
        {\includegraphics[width=0.61\columnwidth]{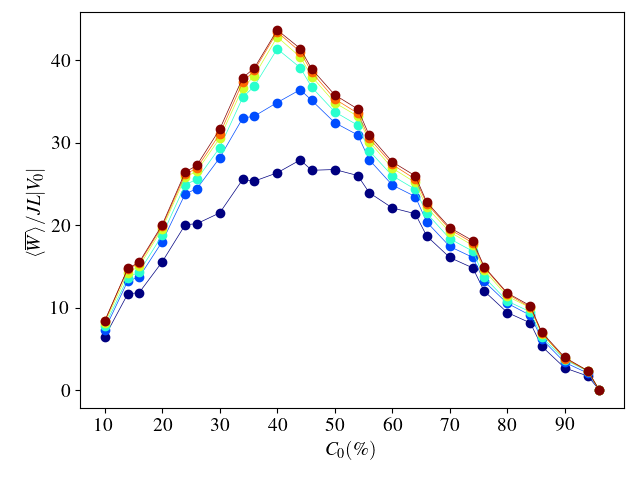}};
    \end{tikzpicture}

    \caption{Panel (a): Single-site entanglement averaged over 100 of the possible impurity configurations versus concentration. Panel (b): $\overline{\braket{W}}/J$, averaged over 100 of the possible impurity configurations,  versus initial concentration
    Panel (c): Average work from panel (b) scaled by the disorder strength $|V_0|$.
    }
    \label{fig:GS_work}
\end{figure*}

Previous results \cite{Canella2019superfluid} for the average ground-state entanglement revealed that the emergence of the full localization in the SIT occurs at the critical concentration $C_C = 100~ n / 2 $ (for attractive disorder). As the disorder strength becomes comparable to the Coulomb attraction, the system starts to exhibit localization and, for $|V|\to\infty$,  it becomes fully localized as the entanglement vanishes. Figure \ref{fig:GS_work}(a) shows the average single-site entanglement $\bar{\mathcal{L}}$ of the initial state\footnote{As we consider sudden quenches, the initial and final state of the system are the same.} as a function of concentration for a chain of $L=100$ sites, average density $n=0.8$,  $U=-5J$ and various values of disorder strength. Figure \ref{fig:GS_work}(b) depicts $\overline{\braket{W}}$,  for the same system undergoing a sudden quench that changes the concentration from  $C_i$ to  $C_{i+1}$.  Our results show that $\overline{\braket{W}}$ for the initial critical concentrations is maximum, corresponding to minimum  entanglement. \ire{This implies that the ground-state is localized at the impurity sites, and of a form very close to eq. (\ref{Psi_C}). For this state, the only non-zero elements contributing to the sum in eq. \ref{eq:average_work+GS}, correspond to impurities which appear in both initial and final impurity configurations,  hence providing up to $N_i/2$ contributions of value $|2V|$.}
This implies that, in finite chains, the average work  is amplified by increasing the disorder strength $V$, both because the value of each contribution is increased and because a stronger impurity increases localization for finite-chains (and finite temperature).
With this in mind, by rescaling the average work by the disorder strength $|V_0|$, we can check when full localization is achieved for this protocol and finite chains. This is explored in  Figure \ref{fig:GS_work}(c), which demonstrates that the rescaled work does not change (full localization for this protocol) for $|V_0|> |U|$.

\begin{figure*}[t!]
\begin{center}
    \begin{tikzpicture}
        \node[label={[font=\normalsize, shift={(-1.75cm,-3mm), align=left}]above:(a) \quad \quad \quad $T=0J/k_B$}] (figa) at (0,0)
        {\includegraphics[ trim={2.75cm 0.5cm 2cm 1.8cm}, clip, width=0.32\textwidth]{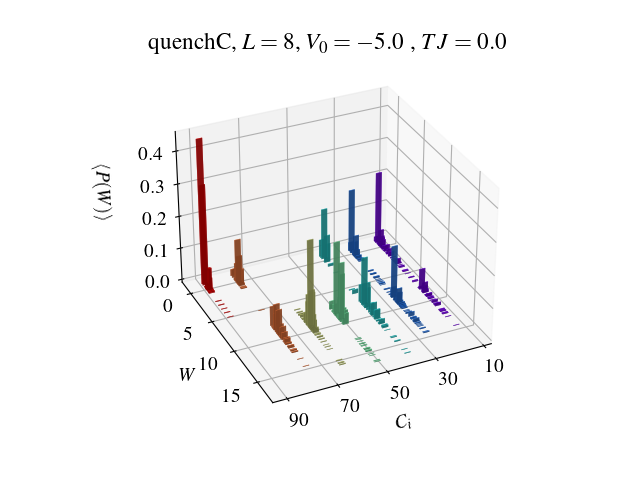}};

        \node[anchor=north west,
       label={[font=\normalsize,
       shift={(-1.75cm,-3mm), align=left}]above:(b) \quad \quad \quad $T=2J/k_B$}
        ] (figb) at ($(figa.north east)+(-0.2cm,0)$)
        {\includegraphics[ trim={2.75cm 0.5cm 2cm 1.8cm}, clip,width=0.32\textwidth]{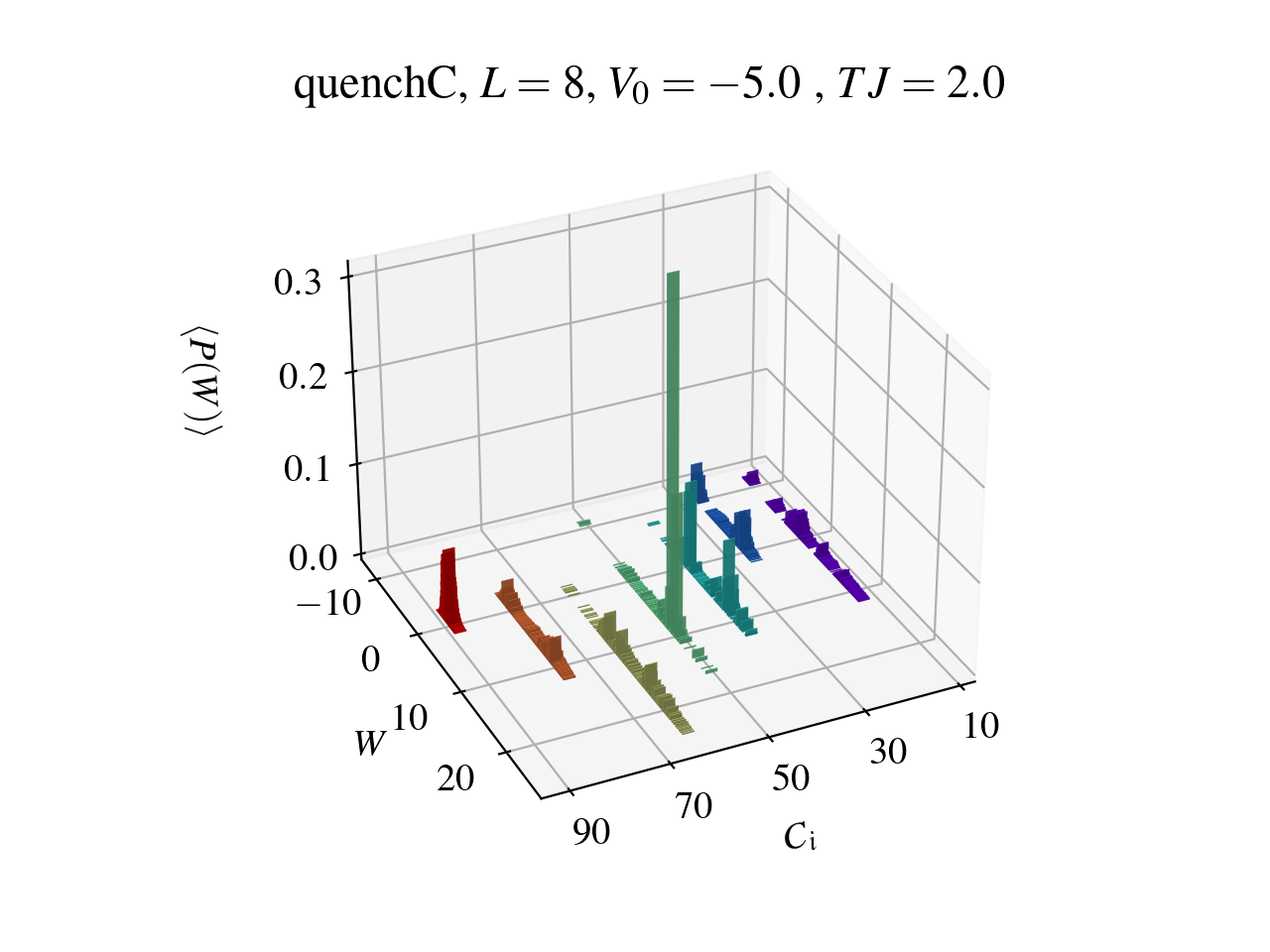}};

        \node[anchor=west,
       label={[font=\normalsize,
       shift={(-1.75cm,-3mm), align=left}]above:(c) \quad \quad \quad $T=30J/k_B$}] (figc) at ($(figb.east)+(0.1cm,0mm)$)
 {\includegraphics[ trim={2.75cm 0.5cm 2cm 1.8cm}, clip,width=0.32\textwidth]{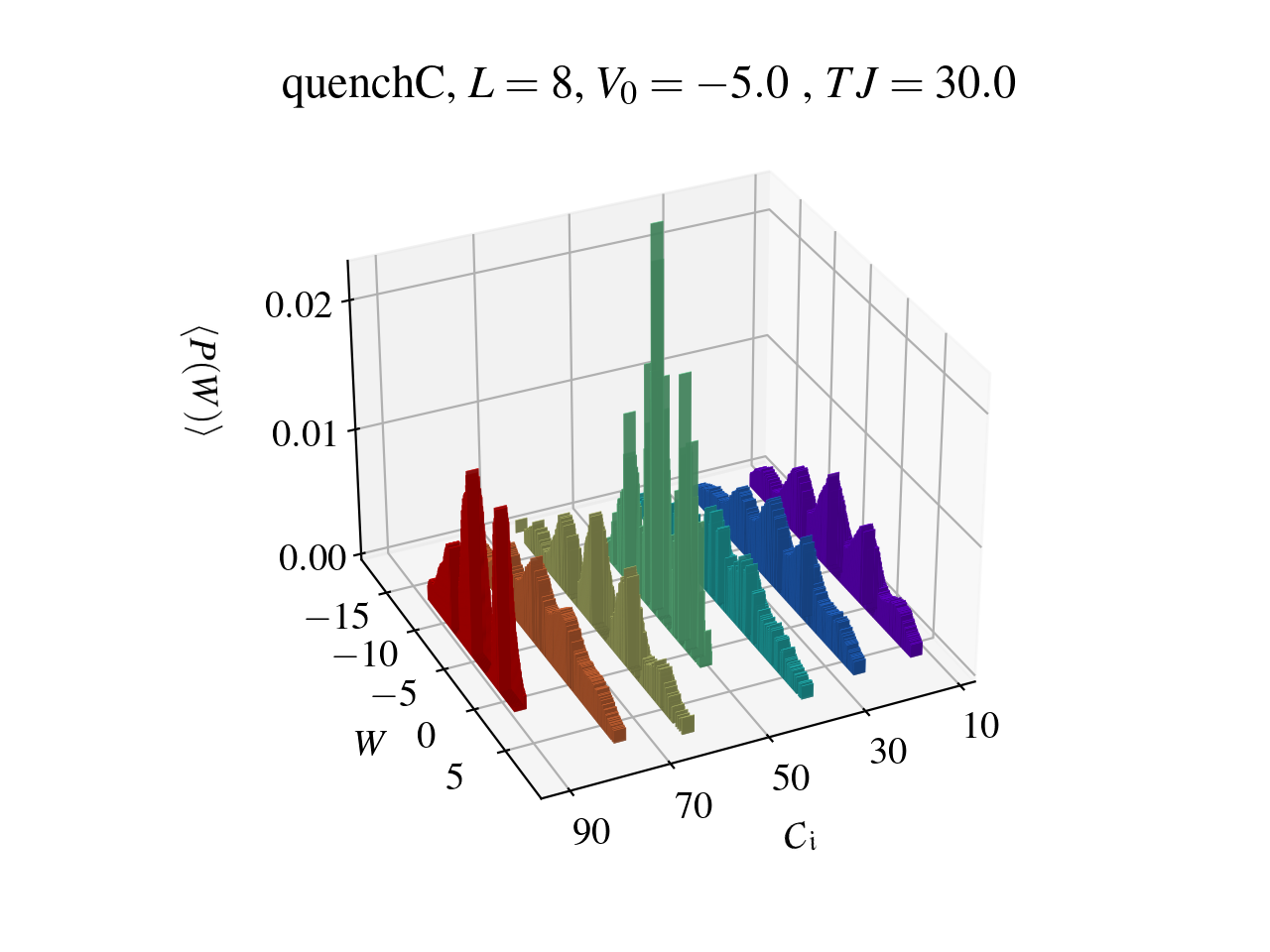}};

    \end{tikzpicture}
\end{center}
\caption{Example of work distributions resulting from quenches in $C$ for chains with $L=8$ sites, with fixed initial and final impurity potential $V_0 = V_f = -5J$ at $T = 0$ (a), $T=2J/k_B$ (b) and $T=30J/k_B$ (c). The initial and final configurations are randomly picked among all possibilities for $C_i$ and $C_{i+1}$.
However, the chosen initial and final configurations' pairs remain the {\it same} at different $T$'s.
}
 \label{fig:work_distribution_vs_Ni_at_T}
\end{figure*}

We now focus our attention on the statistics of the distribution and its changes with the temperature.  Here, we will consider short chains (L=8) at half-filling $n=1$ initially prepared in a thermal state.

Fig. \ref{fig:work_distribution_vs_Ni_at_T} shows examples of $P(W)$ at different concentrations and for the three chosen temperatures,
$T=0J/k_B$ (panel (a)), $T=2J/k_B$ (panel (b)), and $T=30J/k_B$ (panel (c)).
The intermediate temperature has been chosen to be low enough so that the most populated state is the ground-state, while a few low-lying states contribute to the dynamics.

\begin{figure}[htb!]

\begin{center}
	\includegraphics[width=\columnwidth]{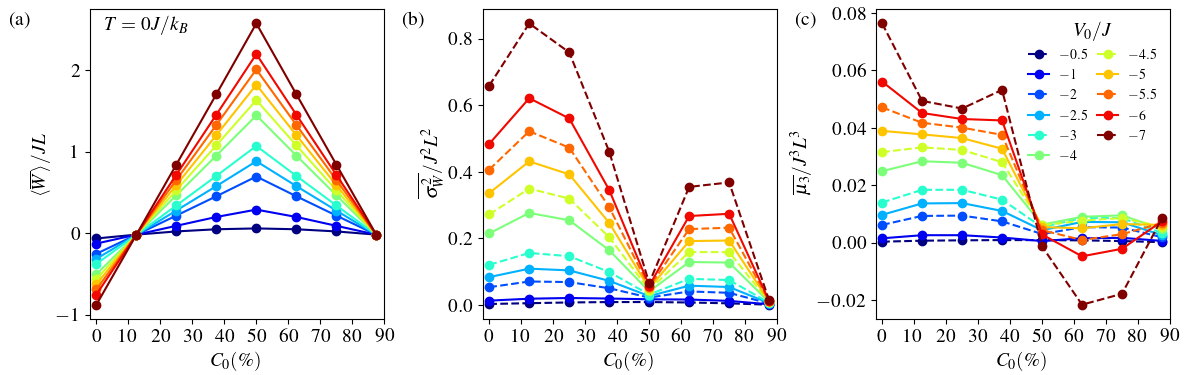}
\end{center}
\begin{center}
	\includegraphics[width=\columnwidth]{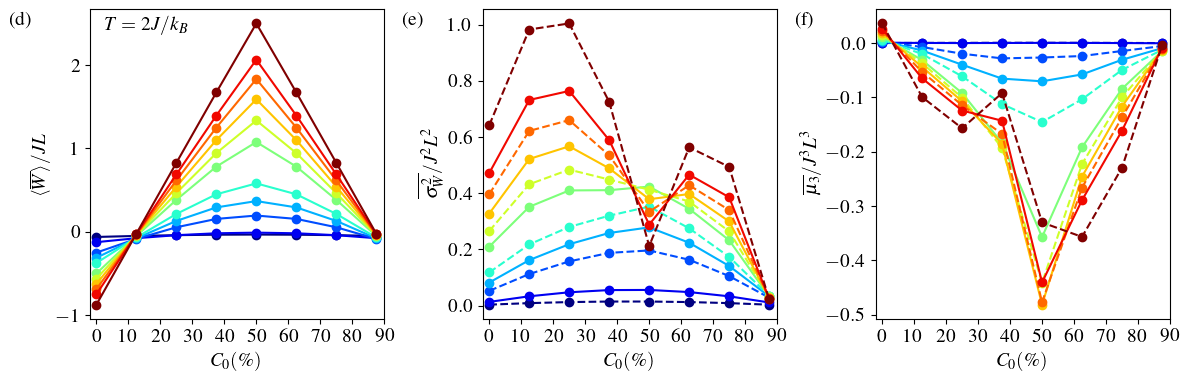}
\end{center}
\begin{center}
	\includegraphics[width=\columnwidth]{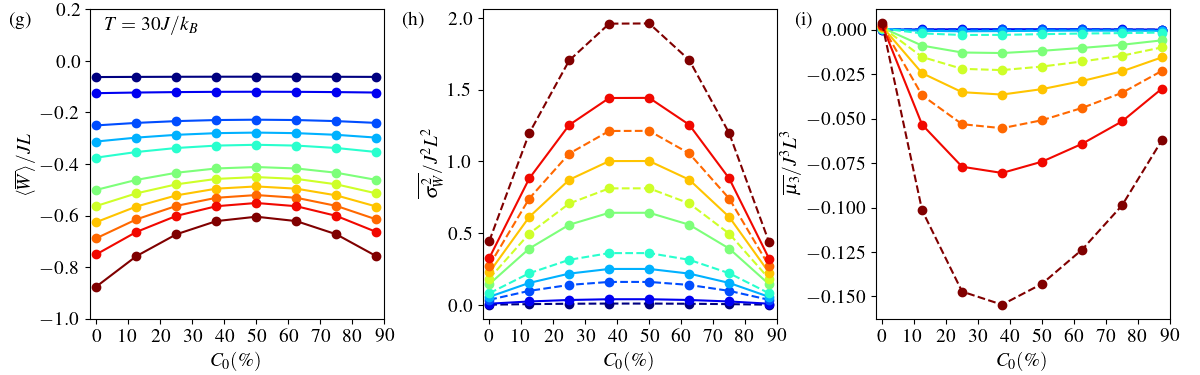}
\end{center}

    \caption{
    Average quantum work (first column), variance (second column) and skewness (third column) versus impurities' concentration of the {\it initial} state for systems with $L=8$ sites, particle interaction $U = -5J$ and for different disorder intensities. The impurities' concentration of the final state is always the next concentration.
   Temperature varies as follows: first row: $ T = 0J/k_B$;  second row: $ T = 2J/k_B$; third  row: $ T = 30J/k_B$;     }

    \label{fig:moments-quenchC-T}
\end{figure}

 Results  for the moments of the work distribution   are shown in Fig. \ref{fig:moments-quenchC-T}, for $T=0J/k_B$, $T=2J/k_B$ and $T=30J/k_B$, as indicated. Here the moments are averaged over all configurations.

 For $T=0J/k_B$ and $T=2J/k_B$ and all impurity potentials,  work can be extracted from the system when starting from zero impurity as initial concentrations.
This is a consequence of eq. \ref{eq:average_work+GS} as it implies that $V_i^0=0$ always, while $V_i^f < 0$ at the impurity site, that will result in negative work.
However this type of contribution competes with positive contributions when starting from nonzero initial impurities concentrations, and work is to be performed on the system in this case - see Fig. \ref{fig:moments-quenchC-T}(a) and (d).

We note that  eq. \ref{eq:average_work+GS} is a consequence of considering a sudden quench dynamics, while finite-time dynamics would allow for redistribution of particle density in response to the new impurity distribution, as observed -- albeit for repulsive impurities and repulsive particle interaction -- in refs \cite{Skelt_2019,Herrera2017_SciRep}. This implies that it cannot be excluded that work could be {\it extracted} from  the system at hand when considering finite-time dynamics.

At $T=0J/k_B$, both variance and skewness show a clear signature of criticality at concentration of $50$ \% see Fig. \ref{fig:moments-quenchC-T} (b) and (c).
The variance almost vanishes, demonstrating, even in a short chain, (almost) no fluctuations for the system at criticality. This is consistent with eq. (\ref{eq:MomentW2_GS}) for a system in the critical state eq. (\ref{Psi_C}), for which  density fluctuations vanish \ire{ at all orders.

Indeed, our numerical results show that also the skewness vanishes at $C_C=50\%$. In addition, it changes its sign from positive, at low concentration, to negative, after $C_C$.} A similar result was observed in Hubbard chains driven by a ramp field tuned  at finite time \cite{QWKrissia}.
Results in \cite{QWKrissia} show this signature in the skewness to be stronger away from sudden quench and towards adiabaticity. In the present work, which used a sudden-quench dynamics, the change-in-sign signature in the skewness appears only for relatively high impurity potentials ($T=0J/k_B$), and it disappears for $T=2J/k_B$ (compare panels (c) and (f) of Fig. \ref{fig:moments-quenchC-T}). At this intermediate temperature, the skewness preserves  only a kink close to the critical concentration value and only at high impurity potentials, while the signature of criticality in the first and second moments is still well marked (Fig. \ref{fig:moments-quenchC-T} (d) and (e)), albeit only at relatively high impurity potentials for the variance. It would be interesting to check if a finite-time dynamics would restore a stronger signature in the third moment, as observed in \cite{QWKrissia}.

Panels (g), (h) and (i) of Fig. \ref{fig:moments-quenchC-T} confirms that any signature of critical behaviour is lost at high temperatures. In particular: work can now be extracted from the system at all concentration values and potentials, fluctuations become maximal around $C_C=50\%$, and the skeweness has a smooth behaviour at all concentrations and impurity potentials.

\subsection{Quench in V: amplifying the disorder strength }
Here we consider quenching a different parameter, the impurities disorder strength.
For each (fixed) impurity concentration and attractive Coulomb interaction of $U=-5J$, the impurities disorder strength will be quenched between an initial value  $V_0$  and the final value $V_f = 2U$. We consider initial disorder strengths from $|V_0| = |U| /10 \ll |U| $ to $|V_0| = 7 \gtrsim |U|$.

\begin{figure}[htb!]

\begin{center}
	\includegraphics[width=\columnwidth]{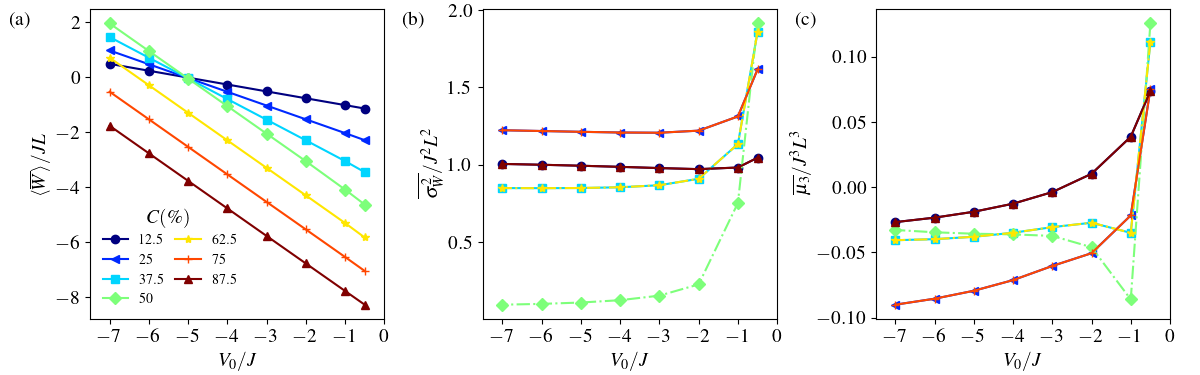}
\end{center}

\begin{center}
	\includegraphics[width=\columnwidth]{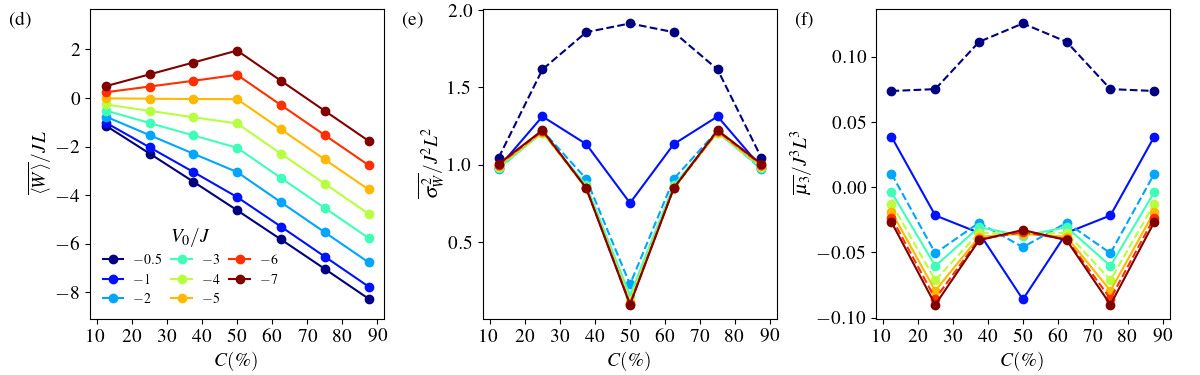}
\end{center}
    \caption{
     First row: Average quantum work (left), variance (middle) and skewness (right) versus $V_0$ (the impurities' strength  of the {\it initial} state) for different disorder concentrations, as labeled. The impurities' strength of the final state is always $V_f = -10J$. Other parameters are: $L=8$, $U = -5J$, and $ T = 0J/k_B$.
   Second row: Same data as for first row but plotted versus the impurities' concentration of the systems considered.  }

    \label{fig:moments-quenchV-T=0}
\end{figure}

Let us first consider the results for $T=0$ when plotted with respect to the varying parameter $V_0$ (Fig. \ref{fig:moments-quenchV-T=0}, upper row).
Comparing mean, variance and skewness, it is difficult to identify consistent signatures of a critical impurity strength. However, the mean (panel (a)) suggests a special role for $V_0=-5J$, at which all curves with impurity concentrations $C\le 50\%$ cross. Indeed, the value $-5J$ is the strength chosen for the Coulomb interaction, so it represents the watershed between either impurity strength or Coulomb attraction being the dominant interaction. Interestingly, a sudden quench in $V$ allows for \ire{work extraction} ($\braket{W}<0$)  for all concentrations when $|V_0| < |U|$ and a decreasing range of concentrations as $V_0$ becomes dominant over the Coulomb interaction.

A trace of this critical role is present in the skewness (panel (b)), where curves close to the critical concentration cross at $V_0\approx -5$. However, the main feature of the skewness is a clear kink at  $V_0=-1$ for intermediate concentrations. In addition, for all concentration and small enough impurity potential the skewness changes sign.
The variance presents some crossovers at small $V_0$ between curves corresponding to different concentrations, but no features at $V_0=U$.

All moments confirm a special role for the data sets collected at the critical concentration $C_C=50\%$ (first row, green lines with diamond markers): the impurity concentration $C$ is identified as a critical parameter even through this protocol.
Hence, in the second row of Fig. \ref{fig:moments-quenchV-T=0}, we plot the same data, but as a function of the impurity concentration $C$ of each calculation. Now all three moments show a clear signature at the critical concentration, all developing a kink at $C=50\%$ as $V_0$ increases. For large enough impurity strength, the variance almost vanishes confirming the quasi-absence of density-density correlations in the corresponding ground state.

The variance and skewness are symmetric with respect to the critical concentration, and present additional features at $C=25\%$ and its symmetric $C=75\%$.
The skewness kinks of panel (c) reflects in panel (f) with an initial potential $V_0<1J$ being too small for the survival of any sign of criticality (panel (f), darkest blue curve), at least for this chain size. This is confirmed by the variance data (panel (e)).
Sensitivity of the skewness to the special value $V_0 = -5J = U$ translates into the extreme at critical concentration  $C=50\%$   changing from a minimum $|V_0| < |U|$ into a maximum $|V_0| > |U|$ in panel (f).  \ire{For increasing $|V_0|$, this maximum tends towards zero skewness, as predicted in Sec. \ref{anal} (not shown).}
In the mean, $V_0 = U$ reflects into a constant $\braket{W}\approx 0$ for $C\le C_C$.

\begin{figure}[htb!]
\begin{center}
	\includegraphics[width=\columnwidth]{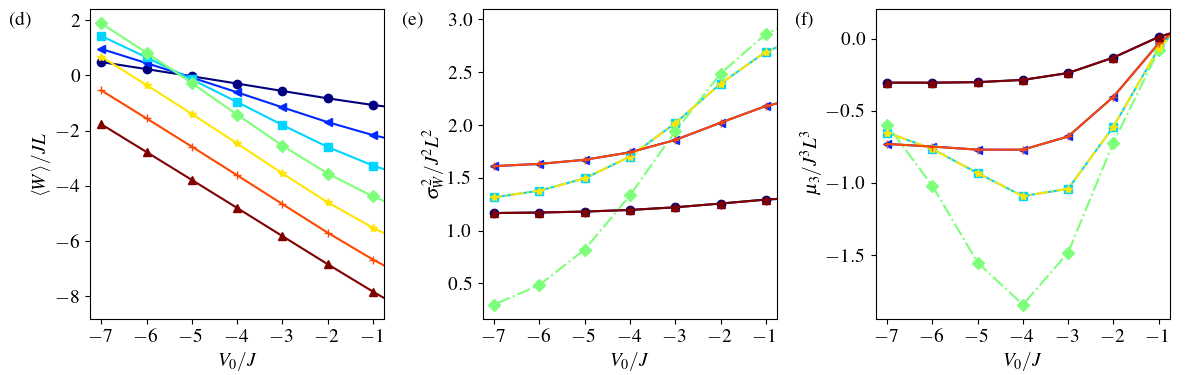}
\end{center}

\begin{center}
	\includegraphics[width=\columnwidth]{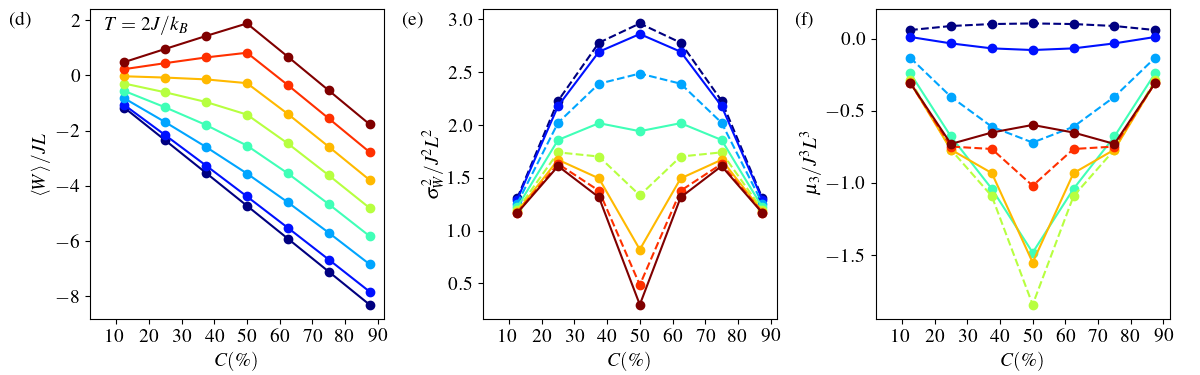}
\end{center}
   \caption{
     Same as Fig. \ref{fig:moments-quenchV-T=0}, but for  $ T = 2J/k_B$.
    }
    \label{fig:moments-quenchV-T=2}
\end{figure}

Increasing the temperature to $T=2 J/k_B$ has little effect in the average work extracted, compare Fig. \ref{fig:moments-quenchV-T=0}(a) and (b) to Fig. \ref{fig:moments-quenchV-T=2}(a) and (b).
However, the second and third moments are more sensitive to the thermal fluctuations, with the signature corresponding to the critical concentration (kinks at $C_C$  in panels (e) and (f))  being washed out for $V_0< 3J$.  Thermal fluctuations also affect the behavior characterizing the watershed value $V_0=U$ in Fig. \ref{fig:moments-quenchV-T=0} (panels (a), (c), and (f)): in Fig. \ref{fig:moments-quenchV-T=2} these signatures are shifted at higher values of $V_0$.
 Thermal fluctuation increase the probability of higher energy transitions, and indeed the work distribution becomes wider (second moment), but also more asymmetric (third moment): compare ranges of the $y$-axis in, e.g., panels (e) and panels (f) of Figs. \ref{fig:moments-quenchV-T=0}  and \ref{fig:moments-quenchV-T=2}.

\begin{figure}[htb!]
\begin{center}
	\includegraphics[width=\columnwidth]{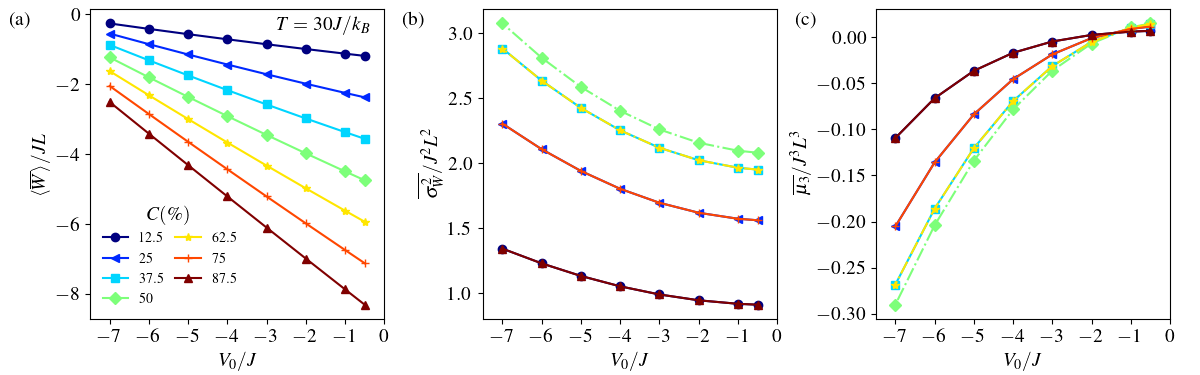}
\end{center}
\begin{center}
	\includegraphics[width=\columnwidth]{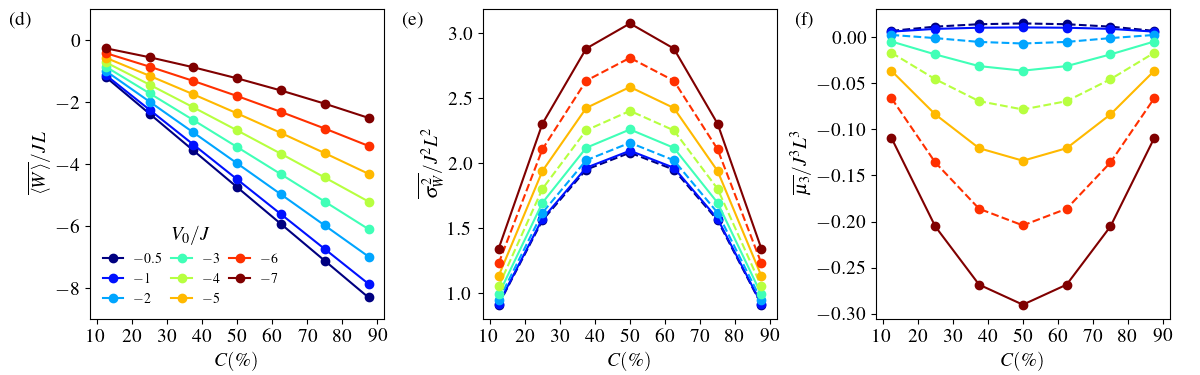}
\end{center}

    \caption{
     Same as Fig. \ref{fig:moments-quenchV-T=0}, but for  $ T = 30J/k_B$.
  }
    \label{fig:moments-quenchV-T=30}
\end{figure}

Finally, high-temperature results ($T=30 J/k_B$, Fig. \ref{fig:moments-quenchV-T=30}) washes away any sign of critical behaviour, as expected. Interestingly, for this lesser quantum-correlated system, work can now be extracted for any parameter combination (i.e. $\hat{\braket {W} < 0}$), suggesting that work has to be done to the system in order to create quantum correlations.

\begin{figure*}[htb!]

\begin{center}
    \begin{tikzpicture}
        \node[label={[font=\normalsize, shift={(-1.75cm,-3mm), align=left}]above:(a) \quad \quad \quad $T=0J/k_B$}] (figa) at (0,0)
        {\includegraphics[ trim={2.95cm 0.5cm 2cm 1.8cm}, clip, width=0.32\textwidth]{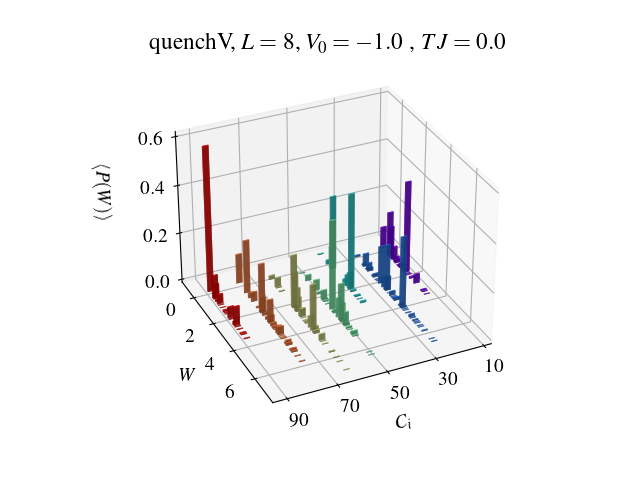}};

        \node[anchor=north west,
       label={[font=\normalsize,
       shift={(-1.75cm,-3mm), align=left}]above:(b) \quad \quad \quad $T=2J/k_B$}
        ] (figb) at ($(figa.north east)+(-0.2cm,0)$)
        {\includegraphics[ trim={2.95cm 0.5cm 2cm 1.8cm}, clip,width=0.32\textwidth]{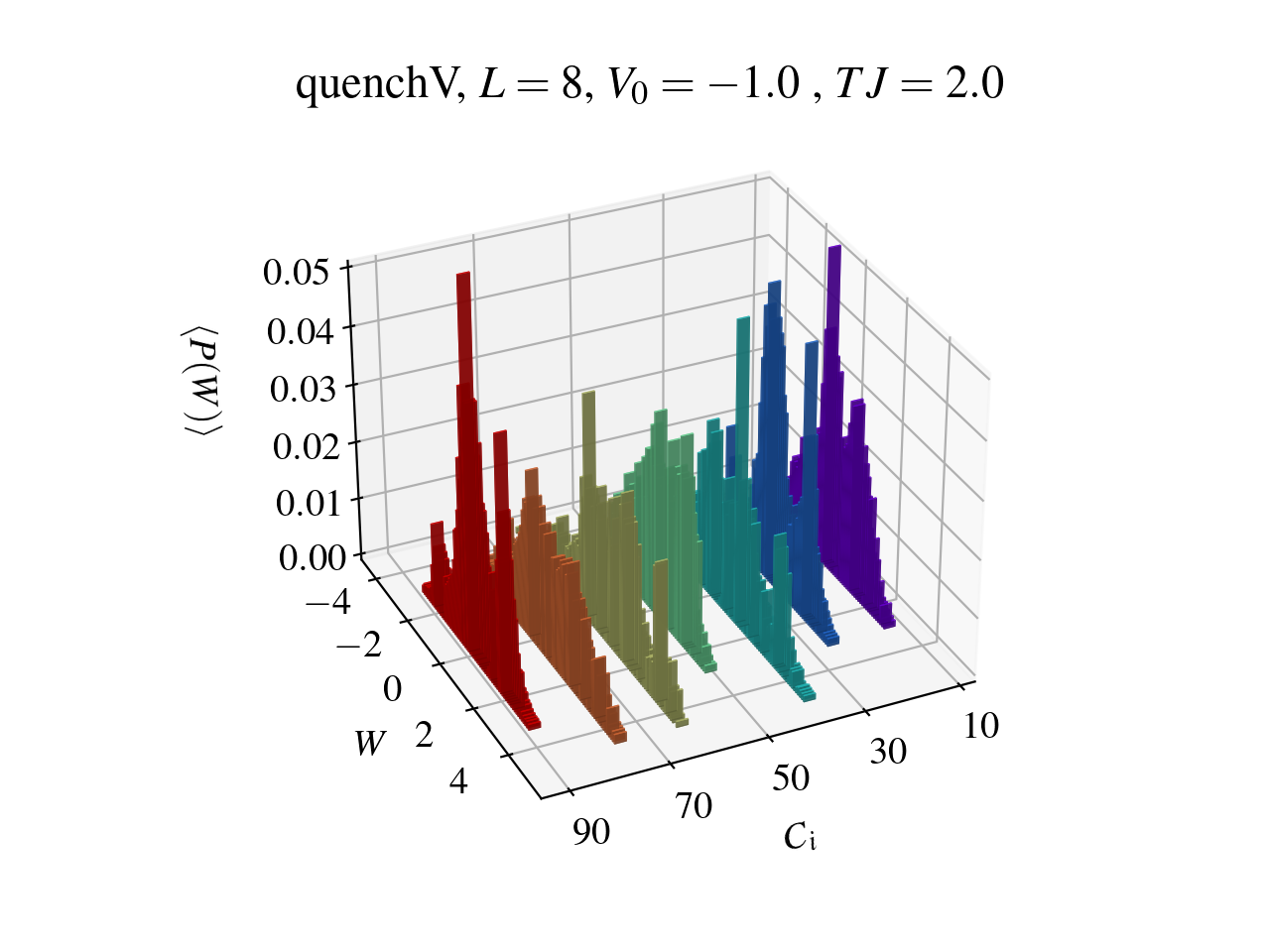}};

        \node[anchor=west,
       label={[font=\normalsize,
       shift={(-1.75cm,-3mm), align=left}]above:(c) \quad \quad \quad $T=30J/k_B$}] (figc) at ($(figb.east)+(0.1cm,0mm)$)
 {\includegraphics[ trim={2.95cm 0.5cm 2cm 1.8cm}, clip,width=0.32\textwidth]{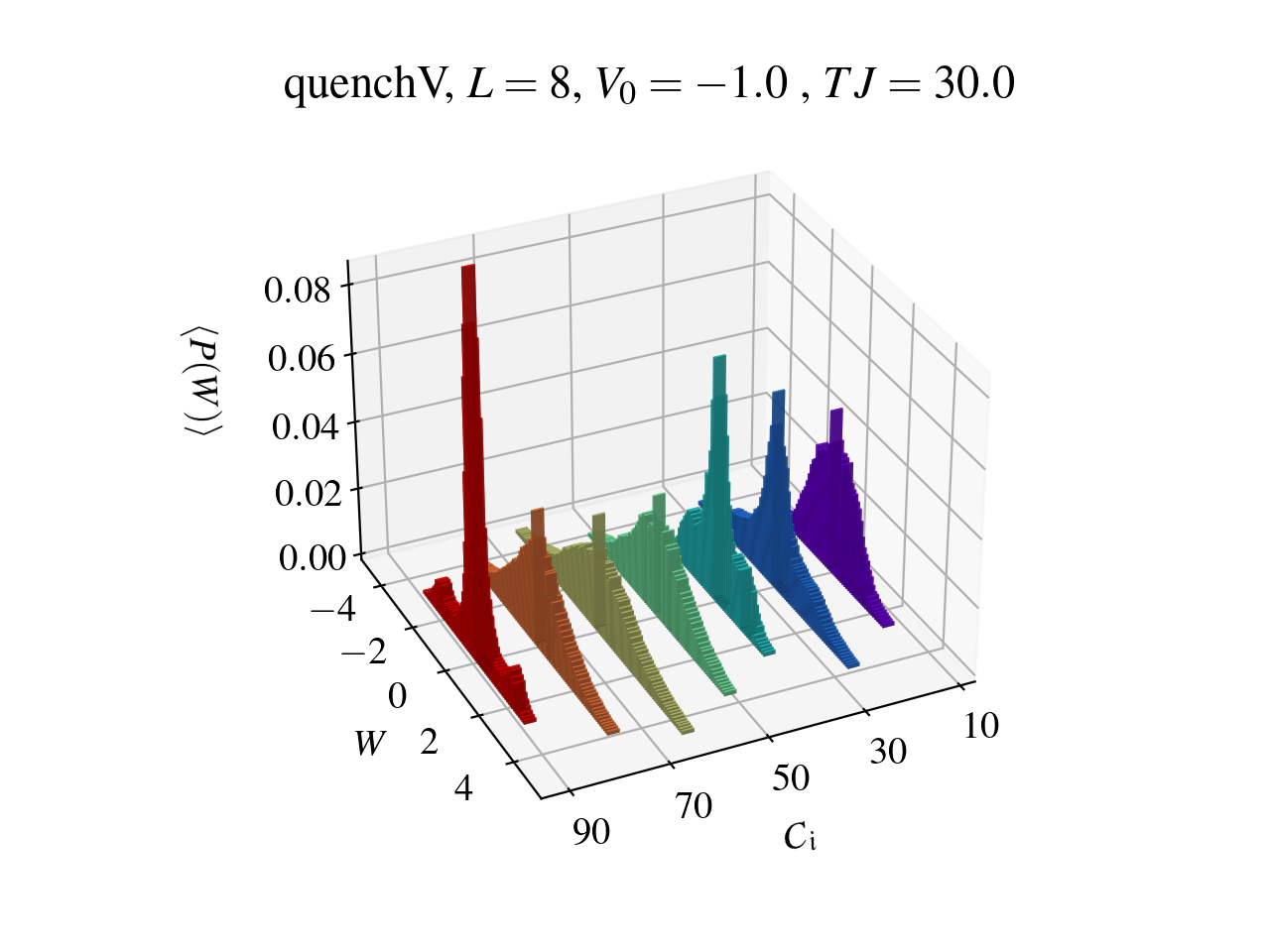}};

    \end{tikzpicture}
\end{center}

\caption{Work distribution for chains with $L=8$ sites resulting from a quench in $V$ from $V_0=-J$ to $V_f=-10J$ with fixed concentration $C_i$ at $T = 0 J/k_B$, $T=2J/k_B$ and $T=30J/k_B$. The initial and final configurations are randomly picked among all possibilities for $C_i$.
However,  initial and final configurations for the {\it same} $C_i$ at different $T$ are the same.
}
 \label{fig:work_distribution_vs_Ni_at_T_quenchV}
\end{figure*}

Figure \ref{fig:work_distribution_vs_Ni_at_T_quenchV} displays one example of $P(W)$ for each temperature. We stress that here we do {\it not} averaged over all possible configurations, but just plot results for a single calculation.
Even so, it can be observed how, with increasing temperature, the distributions at any $C_i$ become wider and then, at $T=30$, more regular (single relevant maximum ) though still displaying a substantial skewness.

\section{Conclusion}
\ire{In this paper, we investigated the statistics of work in a fermionic quantum system undergoing a sudden quench across the superfluid-insulator transition. We demonstrate that at the critical concentration all central moments of the work probability distribution exactly vanish. In fact, in a sudden quench, the work is reduced to a sum of local operators, while the SIT critical state has zero average-site entanglement. This implies that all density fluctuations exactly vanish at all orders and hence all central moments of the work probability distribution follow suit. We note that this property would extend to quenching protocols in other phases for which local entanglement is negligible.

We numerically compared two paths for the SIT in fermionic systems with randomly distributed impurities and explored the first three moments of the work distribution from very low to high temperatures.
Our results indicate that the SIT increases work absorption both when triggered by a change in the impurity concentration or by a quench in  the disorder strength.
Indeed, we demonstrate that the average work is maximized when, at criticality, the single-site entanglement is minimized and explain this analytically. This property could be employed for the implementation of efficient energy storage (quantum batteries).

Work fluctuations become more pronounced away from the critical concentration. Interestingly, deviations from Gaussianity, as revealed by the skewness of the distribution, behave qualitatively differently in the two protocols. When triggering the SIT by varying the impurity concentration the skewness changes sign at the critical point, resembling the behaviour seen in  \cite{QWKrissia}; however the skewness remains negative on both sides of the critical concentration when the transition is triggered by varying the impurities' potential strength.  In this second protocol, both average work and skewness are also highly sensitive to the interplay between impurity and Coulomb potential, with the regime in which the impurity potential dominates been clearly marked in both moments.
We verified that all signatures of the SIT are washed by increasing temperatures.

Future directions include the development of finite-time protocols in which the trade-off between the work and its fluctuations can be controlled dynamically, such as proposed in \cite{QPT-PhysRevLett.120.180605}, and its application to improve the performance of fermionic thermal machines operating at finite temperature.

}

\medskip
\textbf{Acknowledgements} \par 
 KZ acknowledges the
European Research Council Starting Grant ODYSSEY (G.
A. 758403) for financial support and the Northeastern University for computational resources through the Discovery Cluster at the Massachusetts Green High Performance Computing Center (MGHPCC). GC thanks the Department of Physics of the University of York for the kind hospitality and the Coordena\c c\~ao de Aperfeiçoamento de Pessoal de Nivel Superior - Brasil (CAPES) - Finance Code 001.
VVF was supported by FAPESP (2021/06744-8) and CNPq (403890/2021-7; 140854/2021-5). IDA was partly supported by FAPESP (2022/05198-2) and acknowledges the kind hospitality of the Instituto de Física de S\~ao
Carlos, University of S\~ao Paulo, S\~ao Carlos (Brazil).

\medskip

%
\bibliographystyle{MSP}
\bibliography{bib_superfluid}

\end{document}